# In-plane Topological p-n Junction in the

# Three-Dimensional Topological Insulator

# $Bi_{2-x}Sb_xTe_{3-y}Se_y$

*Ngoc Han Tu[1], Yoichi Tanabe[1]\*, Yosuke Satake[1], Khuong Kim Huynh[2], and Katsumi Tanigaki[1,2]\*\**

[1]*Department of Physics, Graduate School of Science, Tohoku University, 6-3 Aramaki, Aoba-ku, Sendai, Miyagi, 980-8577, Japan*

[2]*WPI-Advanced Institute for Materials Research, 2-1-1 Katahira, Aoba-ku, Sendai, Miyagi, 980-8578, Japan*

Email:

\*ytanabe@m.tohoku.ac.jp,

\*\*tanigaki@sspns.phys.tohoku.ac.jp

A topological p-n junction (TPNJ) is an important concept to control spin and charge transport on a surface of three dimensional topological insulators (3D-TIs). Here we report successful fabrication of such TPNJ on a surface of 3D-TI $Bi_{2-x}Sb_xTe_{3-y}Se_y$ thin films and experimental observation of the electrical transport. By tuning the chemical potential of n-type topological Dirac surface of BSTS on its top half by employing tetrafluoro-7,7,8,8-tetracyanoquinodimethane as an organic acceptor molecule, a half surface can be converted to p-type with leaving the other half side as the opposite n-type, and consequently TPNJ can be created. By sweeping the back-gate voltage in the field effect transistor structure, the TPNJ was controlled both on the bottom and the top surfaces. A dramatic change in electrical transport observed at the TPNJ on 3D-TI thin films promises novel spin and charge transport of 3D-TIs for future spintronics.

**Introduction**

Three dimensional topological insulators (3D-TIs) [1,2] provide a new platform for research on electronic properties in materials science. Their helical spin locked Dirac-cone surface states [3,4] are ideal for ultralow dissipative electrical transport, which includes both spin and charge degrees of freedom, and a new design of a logical circuit applicable to various electronic devices can be considered [5-14]. One of the most important features for utilizing such intriguing spin-locked surface states is the topological p-n junction (TPNJ) [7-9]. Theoretically, the inverted helical spin structure in contact between p-type and n-type Dirac cones can bring about superior rectification of spin and charge current. For instance, a significantly large backward scattering of charge current at TPNJ can theoretically give rise to current rectification switching as well as amplification in spin current associated with spin polarization[8]. Such TPNJ studies were recently made for graphene[15-17], where pseudo spin locking is realized via its special geometrical structure and inversion symmetry. Although intensive challenges for fabrication and observation of TPNJs have so far been made also for 3D-TIs as a real spin-locked system [18-22], the important backward scattering and its on-off switching of charge current flow originating from the spin-locked helical Dirac cone surface states have not yet experimentally been reported. Considering the novel electric transport for TPNJ as well as its applications to 3D-TI devices by employing ultralow dissipative spin and charge current, truly firm confirmation and understanding on TPNJ fabricated on the in-plane surface of 3D-TIs are very much warranted.

In this Letter, we report a fabrication of TPNJ on a surface of single crystal 3D-TI $Bi_{1.5}Sb_{0.5}Te_{1.7}Se_{1.3}$ (BSTS)[23-27] ultrathin films. For making the TPNJ without less damage on the surface states, we employ tetrafluoro-7,7,8,8-tetracyanoquinodimethane (F4-TCNQ), an organic molecule known to provide holes to the Dirac fermion surface states of BSTS[28-31] for tuning the chemical potential on its half from n-type to p-type so that a TPNJ can be created on the in-plane surface as shown in Fig.1. Because of both a variety of work functions available from organic semiconductors and a simple and non-damaging fabrication process involved in the present methodology[28-31], tuning of carrier type on the 3D-TI surface by employing an organic semiconductor molecule via interface contact with 3D-TIs is promising for making TPNJs in order to observe the intrinsically emerging electrical transport properties. While the TPNJ is created on the top surface of BSTS thin films, the bottom surface with originally n-type is also modified by the influence of the substrate contact and its chemical potential is further changed under the deposition region of F4-TCNQ. We control the chemical potential of the bottom surface by employing back-gating voltage ($V_G$) in a field effect transistor structure on a $SiO_2$/Si substrate. We clearly observe a ~~step-like~~ dramatic change in electrical transport by changing the $V_G$ and give an experimental clue so far for the highly coveted TPNJ in connection with the theoretical prediction.

**Results**

**Design of topological p-n junction device**

To clarify the electrical transport of TPNJ made on 3D-TIs BSTS, tuning the chemical potential of its topological surface states with keeping the insulating properties of the bulk is essential. BSTS ultrathin films grown on mica followed by transfer to a $SiO_2$/Si substrate can preserve both their high mobility topological surface states and good insulating bulk electronic states [23-27], and are therefore considered to be the best candidate to manipulate the intrinsic electrical transport of TPNJ. Although a tunable Fermi level from n- to p-type in 3D-TIs has been demonstrated for constructing TPNJ[19, 21], intrinsic electric transport experiments for demonstrating TPNJ has not yet been achieved. In order to tune the chemical potential, we employed an organic semiconductor F4-TCNQ, known to be a good electron acceptor, on the BSTS surface. This modification can realize conversion of the n-type topological surface of an as-grown sample with the stoichiometry of $Bi_{1.5}Sb_{0.5}Te_{1.7}Se_{1.3}$ to the p-type one via electron transfer from BSTS to F4-TCNQ owing to its strong electron affinity (5.24 eV [32]) with almost no mechanical and thermal damage thanks to the low temperature process (ca. 60 °C) as well as the $\pi$-orbital soft contact as shown in Fig. 1 (b).

**Electrical transport of topological p-n junction device**

Figure 2(b) showed an evolution of four probe resistance ($R_{xx}$) across the TPNJ made on the top surface between pristine BSTS (n-type surface) and F4-TCNQ/BSTS (p-type surface) as a function of $\mathbf{V}_G$ (also see Supplementary Fig. 1 and 2 and Supplementary Notes 1 and 2 ). In the range of $\mathbf{V}_G$ between -90 and -30 V, where TPNJ states can be created both on the top and the bottom surfaces as explained later based on the detailed electrical transport properties. $R_{xx}$ showed a markedly abrupt and step-like jump at two particular gate voltages: from 12 to 18 k$\Omega$ at $\mathbf{V}_G$ = -30 V and from 14 to 28 k$\Omega$ at $\mathbf{V}_G$ = -90 V. The high resistance values in this window can be compared with greatly lower resistance of ca. 7 k$\Omega$ at the $\mathbf{V}_G$ positively larger than +20 V (n/n contact at the interface both on the top and on the bottom) and ca. 12 k$\Omega$ at the $\mathbf{V}_G$ negatively smaller than -130 V (p/p contact at the interface both on the top and on the bottom), where no TPNJ at the interface between BSTS and F4-TCNQ/BSTS is created. Importantly, no clear difference was experimentally observed between forward and backward bias voltage, being greatly different from the conventional diode properties, reflecting the zero gap linear energy dispersion in 3D-TIs (also See Supplementary Notes 3). A clear enhancement in $R_{xx}$ observed in the TPNJ in the FET structure indicated that the spin-locked topological p-n junction strongly scatters electron transmission, which was theoretically expected.

**Electric transport of pristine BSTS and F4-TCNQ/BSTS device**

In order to clarify the observed TPNJ phenomenon more definitively, useful information was collected on the electrical transport properties for both pristine BSTS and F4-TCNQ/BSTS. Figure 3 (a) shows Hall resistance ($R_{yx}$) of pristine BSTS and F4-TCNQ/BSTS on mica as a function of magnetic field (**B**). Clear experimental evidence was observed, showing that an n-type negative slope (blue color) for pristine BSTS was changed to a p-type positive slope (red color) for F4-TCNQ/BSTS. Longitudinal resistance ($R_{xx}$) of both pristine BSTS (blue color) and F4-TCNQ/BSTS (red color) is also shown in Fig.3 (b) as a function of **B**. A convex-like shaped magnetoresistance ($R_{xx}$) is displayed, showing a typical weak-antilocalization for a spin-momentum locked metallic surface[33]. Importantly, one can clearly see Shubnikov-de Haas (SdH) oscillations superimposed on the $R_{xx}$. After calculations of the longitudinal and the transverse conductance $G_{xx}$ and $G_{xy}$ using $R_{xx}$ and $R_{yx}$ as shown in Supplementary Fig. 3 (a) and (b), the oscillatory parts of $G_{xx}$ ($\Delta G_{xx}$) and $R_{xx}$ ($\Delta R_{xx}$) were separately extracted as a function of **B** as shown in Fig. 3(c) and (d) by subtracting the background using a polynomial function (also see Supplementary Notes 4 ). The amplitude of the oscillations modulated with periodicity of 1/**B** and its decrease in intensity as a function of 1/**B** are consistent with the consequence of the SdH oscillations.

The fact that the oscillation periodicity was clearly different between pristine BSTS and F4-TCNQ/BSTS gives important experimental evidence that the chemical potential was greatly modified by the charge transfer occurring at the interface between BSTS and F4-TCNQ. Via fast Fourier transformation (FFT), two and three oscillation periodicities were estimated to be **B** = 16.7 T and 43.3 T for pristine BSTS and **B** = 3.55 T, 5.95 T and 14.5 T for F4-TCNQ/BSTS, respectively (see Supplementary Fig. 3 and Supplementary

Notes 4). Note that F4-TCNQ molecules make an action as additional scattering centers as well as an electron acceptor on the surface of 3D-TIs whereas the previous report of transition metal phtalocyanine molecules act as an electron donor with absence of backward scattering[34, 35]. As shown in Supplementary Fig. 4 (also see Supplementary Notes 5), pristine BSTS showed a metallic temperature dependence in a wide temperature range whereas the F4-TCNQ/BSTS showed a more insulting-like behavior as well as higher resistance than that of pristine BSTS. These suggest that F4-TCNQ molecules may give additional scattering, resulting in the suppression of the carrier mobility in the F4-TCNQ/BSTS device.

In SdH, the resistance can periodically be modulated by $\Delta G_{xx} \propto cos\left[2\pi\left(\frac{\mathbf{F}}{\mathbf{B}} - \gamma\right)\right]$ (Eq. 1), where $\mathbf{F/B}$ is the oscillation frequency and $\gamma$ is the phase factor. The conventional parabolic energy dispersion gives $\gamma = 1/2$, whereas additional $\alpha = 1/2$ originating from the $\pi$ Berry phase channels [24] is added to $\gamma = 1/2$ in the case of nontrivial topological surfaces. In the Landau level (LL) fan diagram plot, the valley (the peak) in $\Delta G_{xx}$ corresponding to an integer $n$ (a half integer of $n+1/2$) was plotted to deduce further accurate information (Fig. 3 (e) and (f))[36, 37]. Consequently $\alpha \sim 0.5$ ($\mathbf{B} = 16.7$ T and 43.3 T) for pristine BSTS and $\alpha \sim 0.5$ ($\mathbf{B} = 3.55$ T and 14.5 T); $\alpha \sim 0$ ($\mathbf{B} = 5.95$ T) for F4-TCNQ/BSTS (also see Supplementary Fig. 3 and 5 and Supplementary Notes 4). It is noted that trivial carriers ascribed to the surface accumulation layer were also reported in the case of p-type bulk BSTS [21]. Formation of the Rashba-split quantum-well state was also reported at the interface between phthalocyanine and $Bi_2Se_3$ due to the strong band bending and the charge transfer to $Bi_2Se_3$[38]. However, Rashba quantum-well states give the $\pi$ Berry phase[39], and their contribution contradicts with the result of the fan diagram plot and the analysis of the magnetoconductance (see Supplementary Fig. 6 and Supplementary Notes 3).

Due to the influence of a substrate, topological surface states are known not to be identical between the top and the bottom surfaces[40]. For the purpose of having accurate information on both surface states, an FET device with a bottom gate ($\mathbf{V}_G$) was employed. Longitudinal resistance ($R_{xx}$, closed symbols) and Hall coefficient ($R_H$, open symbols) are displayed as a function of $\mathbf{V}_G$ in Fig. 3 (g). The existence of the top and the bottom topological surface is clearly recognized as the two transitions displaying a shoulder around at $\mathbf{V}_G$ = -30 V in both $R_{xx}$ and $R_H$ for F4-TCNQ/BSTS. The nearly $R_H$ = 0 at $\mathbf{V}_G$ = 0 observed for F4-TCNQ/BSTS is the consequence of compensation of the Hall voltage between holes and electrons and indicates that the top surface becomes p-type by the F4-TCNQ modification with leaving the bottom surface still as an n-type, due to the stronger charge transfer from BSTS to F4-TCNQ on the top surface than that on the bottom surface. When positive $\mathbf{V}_G$ is applied, the sign of $R_H$ becomes negative towards a minimum and turns back to zero with keeping its negative sign, indicating that the both surface states can finally become n-type with increasing the number of electron carriers. On the other hand, negative $\mathbf{V}_G$ is employed, the $R_H$ becomes positive first in sign and approaches towards a maximum followed by returning back to a small value with preserving the same sign for both pristine BSTS and F4-TCNQ/BSTS. Accordingly, in the negative $\mathbf{V}_G$ limit, both top and bottom topological surfaces can be tuned to be p-type. The intriguing sign change can be observed in the middle range from $V_G$ = -30 to -90 V straddling from n to p, and this transition can be clearly figured out in the TPNJ experiments in Fig.2b as described earlier.

**Interpretation of resistance switching of TPNJ device**

According to the information on the topological surface states of both top and bottom obtained by electrical transport measurements, we can now deduce an interpretation on the experimental data on our TPNJ experiments as the following scenario shown in Fig. 2 (b). At $V_G$ = 20 V, the first TPNJ is created on the top (p-F4-TCNQ/BSTS and n-BSTS). When $V_G$ = -30 V is applied as the bottom gate potential, the bottom surface of F4-TCNQ/BSTS can be tuned to p-type with leaving the both topological surface of BSTS as n-type. When $V_G$ reaches -90 V, the bottom surface of the pristine BSTS is now tuned to a p-type, while the top surface state of BSTS remains as an n-type with bringing its chemical potential deeper into the Dirac-cone valence band on the surface. Finally, $V_G$ $V_G$ = -130V is applied, all topological surfaces including both the top and the bottom become p-type so that a TPNJ can completely disappear. Consequently, the intriguing TPNJ on the entire surface (both top and bottom) conduction channel can be made only in the range of $V_G$ between -30 and -90 V.

The simple scenario described here seems to explain our experimental observations of the sharp resistance switching in terms of an almost perfect spin-locked backward scattering between the conduction and the valence topological Dirac-cone bands created on the surface at the TPNJ. One may also wonder if energy gap opening on the surface states due to the field induced coupling between the top and the bottom surfaces would be another scenario[41]. However, in the present experiments, we observed a clear resistance jump only at the TPNJ structure and this strongly suggests that the true scenario is TPNJ. Moreover, if the band gap opening could be the reason, one would expect that much higher bulk insulating electric resistivity could appear as the back ground than the present one. One remaining question is whether a sharp resistance jump can be reasonable at the both ends of TPNJ. Another possibility that other conduction channels may be involved in the switch off from the high TPNJ resistance states to the low resistance state should carefully be considered.

Analyses of the magnetoconductance using Hikami-Larkin-Nagaoka formula[33] suggested contribution of additional shallow surface accumulation layers, originating from band bending of the bulk conduction or the valence bands in pristine BSTS and F4-TCNQ/BSTS, when the theoretical interpretations[42,43] are taken into account (see Supplementary Fig. 6 and Supplementary Notes 3). A similar surface accumulation layer was confirmed as a trivial band in the analysis of the SdH oscillations for F4-TCNQ/BSTS as discussed earlier. It should also be noted that a Rashba-like band[44] with $\pi$ Berry-phase cannot be considered because additional conduction channels suggested from the analyses of magnetoconductance are ascribed to weak localization (Berry phase ~ 0)[42,43].

Because exposure to the air and the moisture during the device fabrication process is known to give n-type carriers in the 3D-TIs[24, 45], we assume n-type band bending for pristine BSTS (top and bottom surface) and F4-TCNQ/BSTS (bottom surface), while p-type band bending is considered in F4-TCNQ/BSTS (top surface) as shown in Fig. 4 (a) and (b). In the bottom surface, the negative $V_G$ tunes the chemical potential inside the bulk band gap, resulting in the suppression of the surface accumulation layer in pristine BSTS and F4-TCNQ/BSTS, whereas it continuously tunes the chemical potential of the topological surface states to form the TPNJ shown in Fig. 4(c) and (d). With further decrease in the $V_G$, another surface accumulation layer is induced by the electric field especially at F4-TCNQ/BSTS as shown in Fig. 4 (e) and (f). Because such a conduction channel could give a sudden resistance leakage at the TPNJ leading to a low resistant state due to the large change in density of states of the bulk parabolic band, the surface accumulation layer could switch off the high resistance state of TPNJ to a low resistance at -90 V. A similar situation occurs in the vicinity of the other perfect TPNJ junction limit at $V_G$ = -30 V. On the other hand, in the top surface, because p-type band bending at F4-TCNQ/BSTS continuously exists under negative $V_G$ as shown in Fig. 4 (b), (d) and (f), the p-type surface accumulation layer gives leakage resistance at the TPNJ. Therefore, it would contribute as a little bit broader transition rather than that at $V_G$ = -90 V. It is important to recognize, however, that the sharp switching in resistance occurs exactly at the both junction ends observed at $\mathbf{V}_G$ = -30 and -90 V, where the resistance becomes maximum corresponding to the neutral point of the Dirac surface states as seen in Fig. 3 (g). The scenario described here on the TPNJ in 3D-TIs needs further experimental and theoretical insight in the future.

Since the electronic circuit of TPNJ involving both top and bottom surfaces can be approximated as a parallel circuit connection of resistance, the equivalent electric circuit of the present device is depicted in Fig. 2 (c) and the total resistance expected to be observed experimentally ($R_{obs}$) can be described by

$$1/R_{obs} = 1/[^T R_{BSTS} + {}^T R_{TPNJ} + {}^T R_{F4\text{-}TCNQ/BSTS}] + 1/[^B R_{BSTS} + {}^B R_{TPNJ} + {}^B R_{F4\text{-}TCNQ/BSTS}] \quad (2),$$

where $^T R$ and $^B R$ denote the resistance on the top and the bottom surfaces, respectively. $R_{TPNJ}$ is the resistance at the TPNJ, and $R_{BSTS}$ and $R_{F4\text{-}TCNQ/BSTS}$ are the resistances of the part of pristine BSTS and F4-TCNQ/BSTS. In the F4-TCNQ/BSTS (pristine BSTS) FET, $R_{TPNJ} = R_{BSTS} = 0$ ($R_{TPNJ} = R_{F4\text{-}TCNQ/BSTS} = 0$) is applicable to eq. (2). In the F4-TCNQ/BSTS FET, the Fermi level is different between the top and the bottom surfaces due to the hole transfer from F4-TCNQ. Moreover, the trivial surface accumulation layer existing on the top surface under negative $\mathbf{V}_G$ may supress the ideal on/off ratio. Consequently, eq. (2) gives an asymmetric structure in the $R_{xx}$- $\mathbf{V}_G$ curve. In the case of pristine BSTS, the Fermi level is also different between the top and the bottom surfaces due to the charge transfer from the substrate[46]. In addition, the n-type trivial surface accumulation layer may be suppressed under the negative $V_G$. Therefore, asymmetric structure in $R_{xx}$- $\mathbf{V}_G$ curve with a different on/off switching ratio at the both junction ends is similarly imagined. The features given by eq. (2) are qualitatively consistent with the $R_{xx}$- $\mathbf{V}_G$ curves for both pristine BSTS and F4-TCNQ/BSTS as shown in Fig. 3(g). When TPNJ is created on the both surfaces, high $^T R_{TPNJ}$ and $^B R_{TPNJ}$ can experimentally be expected as long as carrier conduction is prohibited by the spin locking.

According to the recent theoretical calculations[8], small but finitely non-zero transmission probability is predicted when the momentum $\mathbf{P}_y \sim 0$, although the transmission of p-type topological surface band to n-type one, vice versa, is perfectly prohibited for the other directions. Consequently, the resistance at least one order of magnitude larger than that of the normal nn or pp connection is theoretically provided. Our present experimental values (18 - 27 k$\Omega$) are smaller than such expectations (not largely different but one-third to half). One can expect $R_{obs}$= [$^B R_{n\text{-}BSTS}$+$^B R_{n\text{-}F4\text{-}TCNQ/BSTS}$] for $\mathbf{V}_G$= -30 − 20 V, and $R_{obs}$= [$^B R_{p\text{-}BSTS}$+$^B R_{p\text{-}F4\text{-}TCNQ/BSTS}$] for $\mathbf{V}_G$= -130 − -90 V. Our present experiments showed ca. 10 k$\Omega$ of $^B R_{n\text{-}BSTS}$ + $^B R_{n\text{-}F4\text{-}TCNQ/BSTS}$ at $\mathbf{V}_G$ = -30 V and ca. 15 k$\Omega$ of $^B R_{p\text{-}BSTS}$+ $^B R_{p\text{-}F4\text{-}TCNQ/BSTS}$, respectively and they are in good agreement with such considerations.

When one makes more quantitative discussion using the SdH oscillations and the Hall measurements, the following important messages become clear. The SdH oscillations with periodicities of $\mathbf{B}$ = 16.7 T and 43.3 T for pristine BSTS and $\mathbf{B}$ = 3.55 T and 14.5 T for F4-TCNQ/BSTS under $\mathbf{V}_G$ = 0, the carrier concentrations on the Dirac-cone surface bands can be evaluated to be 4.05 ×10$^{11}$ (n) and 1.05×10$^{12}$ cm$^{-2}$ (n) for pristine BSTS and 8.62×10$^{10}$ (p) and 3.52×10$^{11}$ (n) cm$^{-2}$ for F4-TCNQ/BSTS. One can see the clear abrupt jumps in the $R_{xx}$- $\mathbf{V}_G$ in the FET output characteristics at $\mathbf{V}_G$ = -30 and -90 V due to the TPNJ on both top and bottom surfaces. One can also find similar transitions at $\mathbf{V}_G$ = 20 and -130 V, although one transmission channel becomes normal at the bottom surface with either n/n or p/p. These transitions are unfortunately not completely clear for making detailed discussion in the present experiments. By employing the relative dielectric permittivity $\varepsilon_r$ = 3.8 of $SiO_2$, the ideal hole number to be injected to BSTS can be evaluated to be $2.1 \times 10^{12}$ and $6.3 \times 10^{12}$ cm$^{-2}$ by gating of -30 and -90 V, respectively. Considering the fact that $1.05 \times 10^{12}$ (n-type carrier) cm$^{-2}$ of n/p (B) converted to p/p (B) by $\mathbf{V}_G$ = -90 V, and from $3.52 \times 10^{11}$ cm$^{-2}$ of n/n (B) to n/p (B) at $\mathbf{V}_G$ = -30 V, the field effect for carrier accumulation is less by six to seven times than those of the ideal case. This situation seems to be strongly associated with the fact that topological nontrivial Dirac-cone surface states are created via 1:1 correspondence with the bulk and cannot be topologically separated from each other.

**Comparison with previously reported TPNJ device**

As for the comparison with previous report of the vertical and the in-plane TPNJ device[19-22], we didn't observe the threshold-like abrupt increase of the drain current in the present drain voltage ($\mathbf{V}_D$) range between -30 mV and 30 mV [22]. Importantly, no clear difference was experimentally observed between positive and negative $\mathbf{V}_D$, indicating that the switching of current based on the TPNJ scenario is robust for the $\mathbf{V}_D$ range of the present expriments[20,22]. On the other hand in the present device, the resistance jump at TPNJ was suppressed at 300 K ($\sim$ 26 meV) and the potential drop at TPNJ was smaller than that of the gradient induced TPNJ in the $Bi_2Te_3$ single crystal[21]. The $Bi_2Te_3$/$Sb_2Te_3$ heterojunction permitted the minimum film thickness of $Sb_2Te_3$ (15 quintuple layers (QLs)) in order to construct a vertical TPNJ[19], a film with 10 nm (10 QL) in thickness was employed to construct TPNJ in the present F4-TCNQ/BSTS device when $V_G$ ranges from -30 V to 20 V. Therefore, 3D-TI/organic molecule hybrid devices can potentially be utilized also for constructing vertical TPNJ.

**Discussion**

We demonstrated a switching in electrical resistance of the in-plane TPNJ fabricated on a bulk insulating BSTS ultrathin film. By employing F4-TCNQ as a surface modifier, the n-type surface of BSTS thin film was successfully converted to the p-type, and the TPNJ was made on both the top and the bottom surfaces. By employing accurate control of $V_G$ in the FET structure fabricated on the transferred as-grown BSTS ultrathin film to a $SiO_2$/Si substrate, a sharp switching in electrical resistance was observed in four probe resistance measurements. The present experimental results provide important clues for understanding TPNJ in 3D-TIs as well as for logical circuits made on 3D-TI materials in the future.

**Methods**

## Growth and characterization of BSTS thin film

BSTS thin films with 10 nm film in thickness are prepared on mica by van der Waals epitaxy physical vapor deposition[26,2747,48]. We employed the $Bi_{1.5}Sb_{0.5}Te_{1.7}Se_{1.3}$ composition as the source material for the thin film crystal growth, where the surface state is tuned to n-type under the insulating bulk state [23-25]. Since the Dirac neutral point of the surface state in this composition is confirmed between the bulk valence and the conduction bands, the p-n junction on the surface states can be controlled with keeping the bulk insulating. As shown in Fig. 5, a semiconducting-like Sheet resistance ($R_{sheet}$) − temperature (**T**) curve was observed for 50 nm thick BSTS at high temperatures due to the dominant contribution of the insulating bulk electronic states. At low temperatures, it changes to a metallic-like behavior, being consistent with the increase in carrier mobility of the metallic surface states and the decrease in carrier numbers to be thermally populated from the bulk impurity levels. When the thickness is reduced, a semiconducting-like behavior was suppressed even at high temperatures and a metallic behavior became dominant when the thickness reached 10 nm. This can be understood by the fact that the contribution from the bulk states decreased due to the reduction in film thickness and the contribution of the metallic topological surface states became dominant in the total conductance in the present BSTS ultrathin films. By further decreasing the thickness of BSTS, another semiconducting-like behavior again appears in the $R_{sheet}$ − **T** curve for samples with the thickness of 4 and 3 nm. In the 3D-TIs, hybridization of the top and the bottom surfaces is expected to open an energy gap on the topological surface states when the film thickness is sufficiently thin[49]. Based on the experimental facts, the bulk electronic states of the present BSTS thin films are well insulating, and therefore the surface states reasonably give a significant contribution to the electric transport. Film thickness was determined by a surface texture measuring instrument (Surfcom 1400G – Accretech) and an atomic force microscope (AFM, SII NanoTechnology). The quality of the thin film was checked by Raman spectroscopy (Labram HR – 800, Horiba, 628 nm Laser excitation). Raman spectra were acquired from the laser wavelength of 628 nm, and five peaks were observed at 109, 125, 136, 156, and 163 $cm^{-1}$, which are consistent with the previous results [26, 27].

**Device fabrication**

For fabricating a back gate FET device, BSTS thin films are peeled from a mica substrate using deionized water and transferred onto a $SiO_2$(300 nm)/Si substrate [27]. An electrical contact was prepared by a thermal evaporation method using a metal mask (Au: 50 nm) and connect by 25 μm gold wire using silver paste (Dupont4922). F4-TCNQ was deposited on the half of a BSTS thin film by thermal evaporation under the pressure of $10^{-6}$ Pa to form the p-n junction on the top surface of BSTS thin films. Electrical transport measurements were performed by a standard four probe method using a semiconducting devise analyzer (Agilent B1500), a nano-voltmeter (Agilent 34420A), and a signal generator (Keithley 6221) under magnetic fields below 9 T in the temperature range of 2 to 300 K.

**Data availability**

The data that support the findings of this study are available from the corresponding author upon reasonable request.

**Acknowledgements**

We are grateful to H. Shimotani and T. Kanagasekaran to support the usage of an organic evapolator. We also grateful to S. Ikeda, R. Kumashiro and K. Saito (Common Equipment Unit of Advanced Institute for Materials Research, Tohoku University) to support various experiments. K. Nomura, K. Sato, R. Nakai and S. Heguri were acknowledged for fruitful discussion on TPNJ. This work was sponsored by the fusion research funds of "World Premier International (WPI) Research Center Initiative for Atoms, Molecules and Materials", MEXT, Japan; JSPS KAKENHI Grant Number JP15K05148.

**Author contribution**

Y.T, K.T. designed the project. N.H.T, Y.T. Y.S. synthesized $Bi_{2-x}Sb_xTe_{3-y}Se_y$ (BSTS) thin film by the van deer Waals epitaxy and fabricated BSTS p-n junction device. N.H.T., Y.T. characterized BSTS thin films. N.H.T., Y.T., K.K.H. measured BSTS p-n junction devices. N.H.T., Y.T., K.T. wrote a manuscript. All authors discussed the results and commented on the manuscript.

**Additional information**

Correspondence and requests for materials should be addressed to Y.T. and K.T.

**Competing financial interests**

The authors declare no competing financial interests.

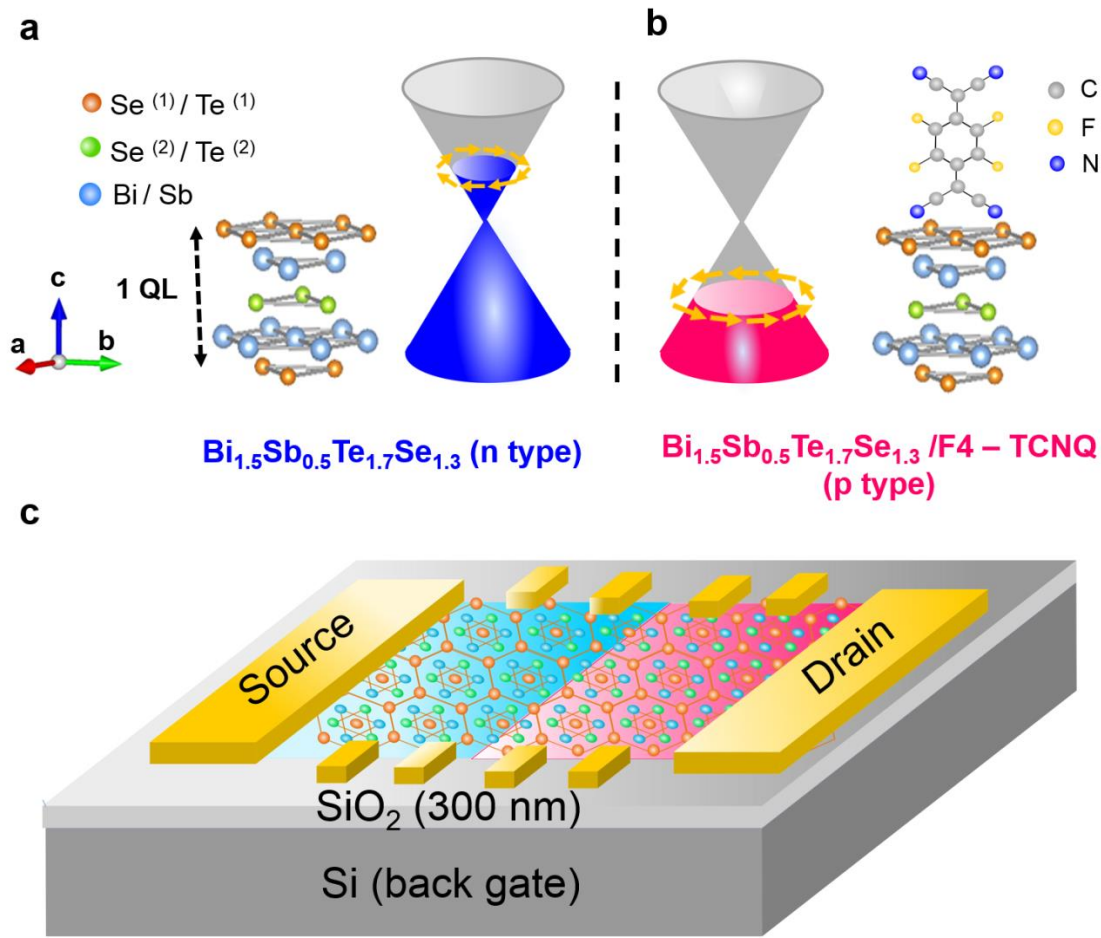

**Figure 1| Schematic view of TPNJ device.** (a), (b) electronic structure of BSTS and F4-TCNQ/BSTS and (c) topological p-n junction device. In order to make a local carrier tuning on the 3D-TI thin film, F4-TCNQ molecules are partly deposited on the n-type BSTS thin film surface.

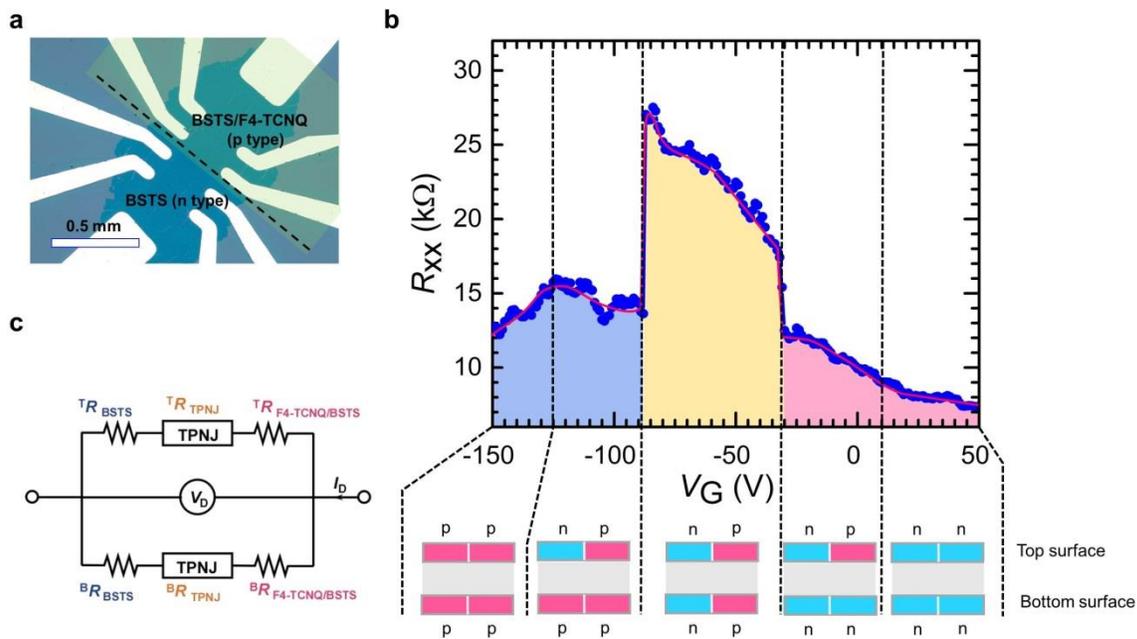

**Figure 2| Electric transport of TPNJ device.** (a) Microscope view of a TPNJ device. BSTS thin film with 10 nm thickness grown on mica was transferred on a SiO₂/Si substrate to fabricate a bottom gate field effect transistor. F4-TCNQ was partly deposited on the BSTS thin films to realize injection of holes on the partial area of BSTS. (b) Gate voltage (**V**$_G$) dependence of $R_{xx}$ across the TPNJ made between pristine BSTS and F4-TCNQ/BSTS at 2 K. Schematic views of the TPNJ for the top and the bottom surface states available from the present device are illustrated. (c) The equivalent electric circuit of the present device. The electronic circuit of TPNJ involving both top and bottom surfaces can be described as a parallel circuit connection of resistance.

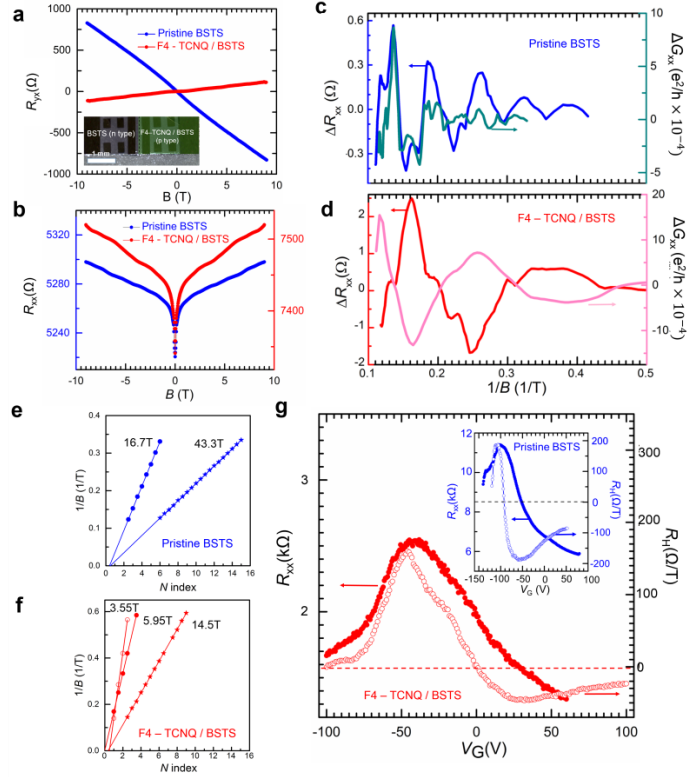

**Figure 3| Electric transport of BSTS and F4-TCNQ/BSTS device.** (a), (b), Magnetic field (**B**) dependence of transverse and longitudinal resistance ($R_{yx}$ and $R_{xx}$) at 2 K for pristine BSTS and F4-TCNQ/BSTS on mica. Inset of (a) shows a microscope view of pristine BSTS and F4-TCNQ/BSTS device. (c), (d) 1/**B** dependence of the oscillatory parts of longitudinal resistance ($\Delta R_{xx}$) and longitudinal conductance ($\Delta G_{xx}$) at 2 K. (e), (f) A Landau level fan diagram plot, where the valley (the peak) in $\Delta G_{xx}$ indicates as an integer $n$ (a half integer $n + 1/2$ ). (g), Gate voltage (**V**$_G$) dependence of $R_{xx}$ and Hall coefficient ($R_H$) calculated from $R_{yx}$ at **B** = ± 9 T for a F4-TCNQ/BSTS bottom gate field effect transistor (FET) prepared on a SiO$_2$/Si substrate. Inset shows **V**$_G$ dependence of $R_{xx}$ and $R_H$ for a device for pristine BSTS.

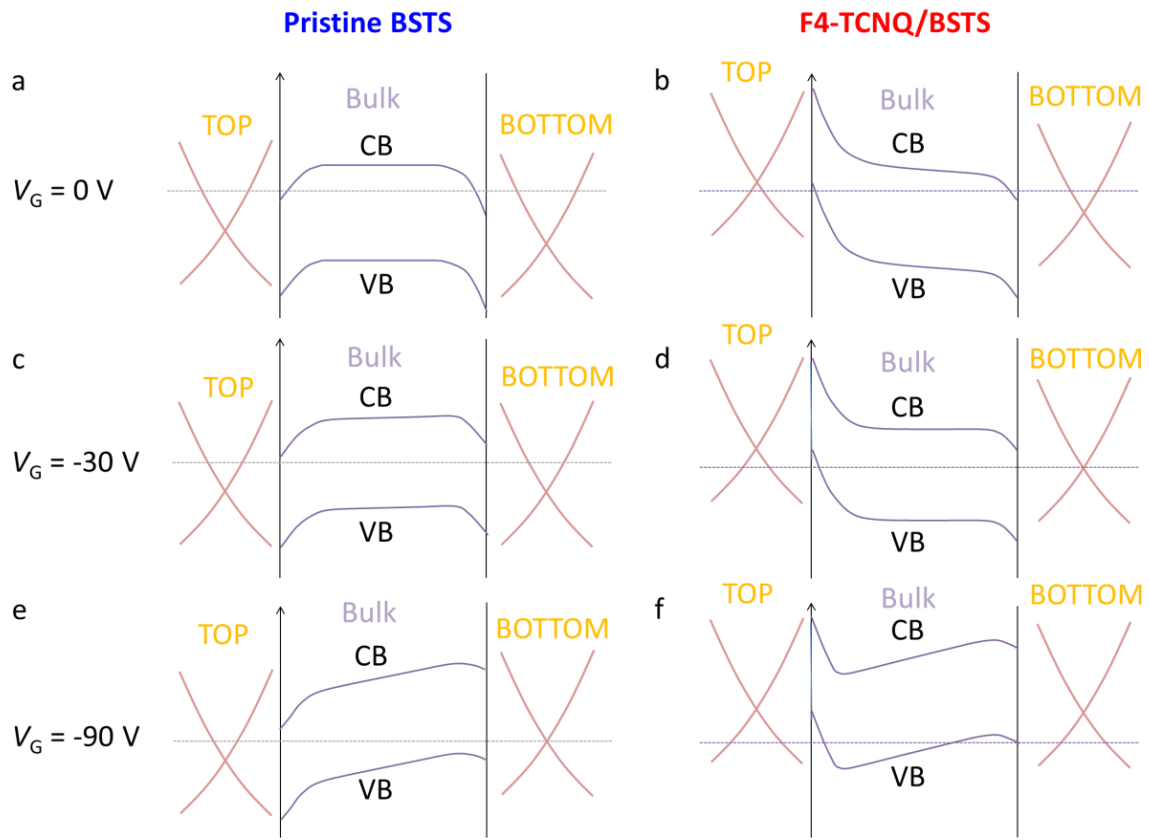

**Figure 4| Electronic structure of the present TPNJ device.** The electronic structure of pristine BSTS and F4-TCNQ/BSTS in a topological p-n junction device under (a), (b) $\mathbf{V}_G$ = 0 V, (c), (d) $\mathbf{V}_G$ = -30 V, and (e), (f) $\mathbf{V}_G$ = -90 V. A high resistance states of TPNJ in the bottom surface can be switched on/off by tuning the chemical potential of the trivial surface accumulation layer.

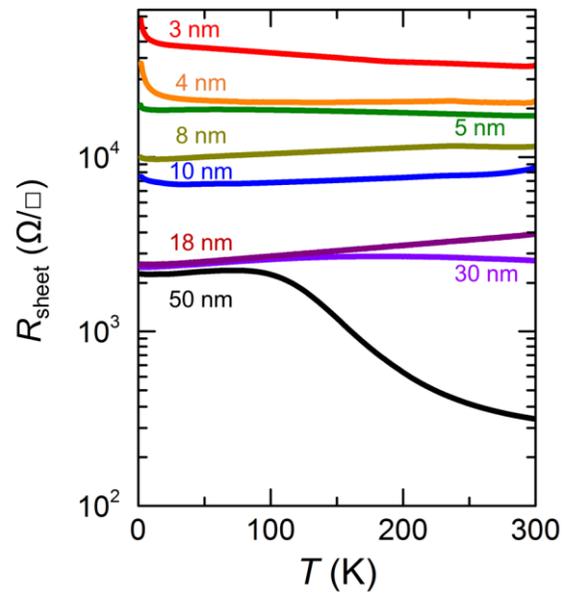

**Figure 5| Thickness dependence of electrical transport for BSTS thin films.** Temperature dependence of the electrical transport for $Bi_{1.5}Sb_{0.5}Te_{1.7}Se_{1.3}$ thin film with various film thickness grown by the van der Waals epitaxy physical vapor deposition are displayed.